\documentclass[aps,prl,twocolumn,superscriptaddress,showpacs,amsmath,amssymb,10pt]{revtex4-1}

\usepackage{graphicx}

\usepackage{siunitx}
\DeclareSIUnit\stiffness{\pico\newton/\micro\meter}
\DeclareSIUnit\kbT{k_{\text{B}}T}
\usepackage{xcolor}

\newcommand{\resub}[1]{\textcolor{black}{#1}}

\definecolor{linkcolor}{rgb}{0,0,0.6} 
\usepackage[pdftex,colorlinks=true, pdfstartview=FitV, linkcolor= linkcolor, citecolor= linkcolor, urlcolor= linkcolor, hyperindex=true,hyperfigures=true]{hyperref} 

\begin{document}


\title{Stationary and transient Fluctuation Theorems for effective heat flux between hydrodynamically coupled particles in optical traps}


\author{A. B\'{e}rut}
\email[]{antoine.berut@ens-lyon.org}
\affiliation{Universit\'{e} de Lyon \\
Laboratoire de Physique, \'{E}cole Normale Sup\'{e}rieure de Lyon (CNRS UMR5672), 46 All\'{e}e d'Italie 69364 Lyon Cedex 07, France}
\author{A. Imparato}
\affiliation{Department of Physics and Astronomy, Aarhus University, DK-8000 Aarhus C, Denmark}
\author{A. Petrosyan}
\author{S. Ciliberto}
\affiliation{Universit\'{e} de Lyon \\
Laboratoire de Physique, \'{E}cole Normale Sup\'{e}rieure de Lyon (CNRS UMR5672), 46 All\'{e}e d'Italie 69364 Lyon Cedex 07, France}


\date{December 3, 2015}

\begin{abstract}
We experimentally study the statistical properties of the energy fluxes between two trapped Brownian particles, interacting through dissipative hydrodynamic coupling, submitted to an effective temperature difference $\Delta T$, obtained by random forcing the position of one trap. We identify effective heat fluxes between the two particles and show that they satisfy an exchange fluctuation theorem (xFT) in the stationary state. We also show that after the sudden application of a temperature gradient $\Delta T$, \resub{the total} hot-cold flux satisfies \resub{a} transient xFT for any integration time whereas \resub{the total} cold-hot flux only does it asymptotically for long times.
\end{abstract}

\pacs{}

\maketitle


Nowadays the energetics of small devices, as for example nano-motors, is a widely studied problem which is important not only from a fundamental point of view but also for applications. In these small systems the energies involved are of the order of few \si{\kbT} and the statistical properties of their fluctuations cannot be neglected~\cite{Seifert2012,Ciliberto_ann}. Experimentally these statistical properties have been widely studied in systems in contact with a single heat bath~\cite{Ciliberto_ann}. 
Conversely, energy fluxes between systems kept at different temperatures have been analyzed, within the framework of stochastic thermodynamics, only in a few experiments in electronic circuits~\cite{Ciliberto2013,Ciliberto2013a} and in single electron-boxes~\cite{Koski2013}. Moreover these kinds of energy fluxes have been theoretically studied only in systems with a conservative coupling~\cite{Visco2006,Saito07,Fogedby11a,Saito11,Fogedby12,Fogedby14}. Thus the question of the possible  modifications of their statistical properties when the coupling is dissipative has never been addressed.

In this letter we analyze this question in an experiment where two trapped Brownian particles are viscously coupled and submitted to an effective temperature difference obtained by randomly displacing the position of one of the two traps. We also study to which extent the energy exchanged between the two particles can be considered as a real heat flux and the random forcing as a real heat bath. Indeed we find that energy fluxes satisfy a stationary exchange fluctuation theorem (xFT):
\begin{equation}
\ln \left( \frac{P(Q_{\tau})}{P(-Q_{\tau})} \right) \underset{\tau \to \infty}{=} \frac{1}{k_{\text{B}} }
\left(  \frac{1}{T_1}-\frac{1}{ T_2} \right) Q_{\tau}
\end{equation}
where \resub{$k_{\mathrm{B}}$ is the Boltzmann constant, and} $P(Q_{\tau})$ is the probability that an amount of (effective) heat $Q_{\tau}$ is exchanged during a time $\tau$ between the two systems at (effective) temperatures $T_1$ and $T_2$.
Furthermore, during the sudden application of the temperature gradient, \resub{the total} hot-cold flux satisfies the transient xFT for any integration time whereas \resub{the total} cold-hot flux only does it asymptotically for long times. This asymmetric behavior has been predicted for systems with a conservative coupling and we extend it here to the case of viscous coupling. We also show that it is only possible to recognize the dissipative nature of the coupling in the case when the two traps have a different stiffness, i.e. the system is asymmetric.

{\it Experimental set-up}. 
\resub{The experiment is performed using a set-up which is similar to the one described in~\cite{Berut2014}:} a custom-built vertical optical tweezers with an oil-immersion objective (HCX PL. APO $63\times$/$0.6$-$1.4$) focuses a laser beam (wavelength \SI{532}{\nano\meter}) to creates a quadratic potential well where a silica bead (radius $R = \SI{1}{\micro\meter} \pm 5\%$) can be trapped. The beam goes through an acousto-optic deflector (AOD) that allows to modify the position of the trap very rapidly (up to \SI{1}{\mega\hertz}). By switching the trap at \SI{10}{\kilo\hertz} between two positions we create two independent traps, which allows us to hold two beads separated by a fixed distance. The beads are dispersed in bidistilled water at low concentration to avoid interactions with multiple other beads. The solution of beads is contained in a disk-shaped cell (\SI{18}{\milli\meter} in diameter, $\SI{1}{\milli\meter}$ in depth). The beads are trapped at \SI{15}{\micro\meter} above the bottom surface of the cell. The position of the beads is tracked by a fast camera with a resolution of \SI{115}{\nano\meter} per pixel, which after treatment~\cite{Crocker1996} gives the position with an accuracy greater than \SI{5}{\nano\meter}. The trajectories of the bead are sampled at \SI{800}{\hertz}. The stiffness of the traps $k$ (typically about \SI{4}{\pico\newton/\micro\meter}) is proportional to the laser intensity and can be modified by adding neutral density filters or by changing the time that the laser spend on each trap. The two particles are trapped on a line (called ``x axis'') and separated by a distance $d$ which is tunable. For a distance of a few radiuses (typically \SI{4}{\micro\meter}) the Coulombian interaction between the particle surfaces is negligible.

\resub{The ``effective temperature'' of one of the two  particles (for example particle $1$) is obtained by sending a Gaussian white noise (filtered at \SI{1}{\kilo\hertz}) to the AOD; in such a way that the position of the corresponding trap is moved randomly along the direction where the particles are aligned (x-axis). If the amplitude of the displacement is sufficiently small to stay in the linear regime it creates a random force on the particle which does not affect the stiffness of the trap~\cite{Berut2014}. When the random force is switched on, the bead quickly reaches a stationary state with an effective temperature for the randomly forced degree of freedom~\cite{Martinez2013,Berut2014}.}

{\it Hydrodynamic coupling model}. 
\resub{The two particles interact only  through the motion of their (viscous) surrounding fluid. This hydrodynamic coupling, in low Reynolds-number flow, can be described by a mobility matrix $\mathcal{H}$ linking the particles velocities to the forces acting on them~\cite{Meiners1999,Bartlett2001,Ou-Yang2002}. This hydrodynamic model was already used in non-equilibrium situations with a shear-flow~\cite{Zimmermann2009} and we have already shown in~\cite{Berut2014} that its predictions are in good agreement with experimental observations when one of the two particles is randomly forced.}

\resub{For two identical particles of radius $R$ trapped at positions separated by a distance $d$ sufficiently larger than their typical displacements, $\mathcal{H}$ is the Rotne-Prager diffusion tensor~\cite{Herrera2013}:
\begin{equation}
\mathcal{H} = 
\begin{pmatrix} 1/\gamma & \epsilon/\gamma \\
 \epsilon/\gamma & 1/\gamma \end{pmatrix}
\end{equation}
where $\gamma$ is the Stokes friction coefficient ($\gamma = 6 \pi R \eta$ where $\eta$ is the viscosity of water) and $\epsilon = \frac{3R}{2d} - \left(\frac{R}{d}\right)^3$ is the coupling coefficient.\newline
In our case, the particle $2$ is in contact with a thermal bath at room temperature $T_2=T$ and the particle $1$ is kept at an effective temperature $T_1 = T + \Delta T$. It follows that the longitudinal motion of the two thermally excited trapped particles is described by two coupled overdamped Langevin equations~\citep{Berut2014}:
\begin{equation}
\left\{
  \begin{array}{l}
    \gamma \dot{x}_1 = -k_1 x_1 - \epsilon k_2 x_2  + \xi_1 \\
    \gamma \dot{x}_2 = -k_2 x_2 - \epsilon k_1 x_1 + \xi_2
  \end{array}
\right.
\label{rforce:eq:coupledLangevinsimplified}
\end{equation}
where $x_i$ is the position of the particle $i$ relative to its trapping position, $\dot{x}_i$ is the time derivative of $x_i$, and $\xi_i$ are the equivalent random forcing. The equivalent random forcing are given by: 
\begin{equation}
  \begin{array}{rcl}
    \xi_1 & = &  f_1 + \epsilon f_2 + f^*\\
    \xi_2 & = & f_2 + \epsilon f_1 + \epsilon f^*
  \end{array}
\end{equation}
where the $f_i$ are the equilibrium Brownian random forces of the bath at temperature $T$, and $f^*$ is the external random force added on particle 1, that is characterized by the effective temperature $\Delta T$. The random forces are all zero on average, and verify:
\begin{equation}
	\begin{array}{ccl}
		\langle f_i(t) f_j(t^{\prime}) \rangle &=& 2k_{\mathrm{B}}T \, (\mathcal{H}^{-1})_{ij} \, \delta(t-t^{\prime}) \\ 
		\langle f^*(t)f^*(t^{\prime}) \rangle &=& 2k_{\mathrm{B}} \Delta T \gamma \delta(t-t^{\prime}) \\
		\langle f^*(t) f_i(t^{\prime}) \rangle &=& 0.
	\end{array}
\end{equation}}

\resub{The system of equations~(\ref{rforce:eq:coupledLangevinsimplified}) shows the non-conservative nature of the hydrodynamic coupling. Indeed, in the general case where $k_1 \neq k_2$, the coupling terms $-\epsilon k_i x_i$ cannot be written as the partial derivatives of a single potential $U$ with respect to $x_1$ and $x_2$, respectively. This is a very important difference between our system and those of references~\cite{Ciliberto2013,Ciliberto2013a}, which have a conservative coupling.} These systems can be described with the equivalent Fokker-Planck formalism, as discussed in~\cite{Fogedby12} for conservative forces, and in~\cite{NewPap} for the present case with non-conservative interactions.

\resub{Experimentally}, the values of $k_1$ and $k_2$ can be calibrated beforehand with usual methods~\cite{BergSorensenRSI2004}. The values of $\Delta T$ and $\epsilon$ can be computed from the values of the variances of $x_1$ and $x_2$~\cite{Berut2014}. Note however that when the two traps are created by a single laser switched rapidly with an AOD, the values of $\epsilon$ that are measured are always smaller ($\sim \SI{25}{\percent}$) than the Rotne-Prager predictions (unlike what is observed when the two traps are created with two static crossed polarized beams, as in~\citep{Berut2014}). Thus, the value of $\epsilon$ needs to be measured each time.

\begin{figure}[!ht]
\includegraphics[width=0.9\linewidth]{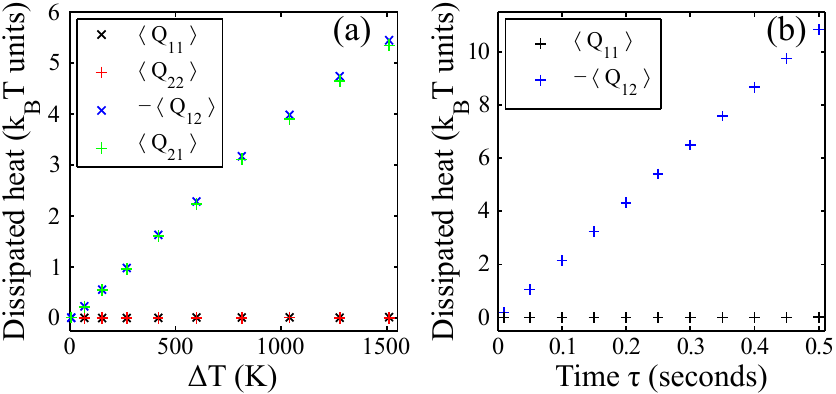}
\caption{\textbf{(a)} Average value of the four $Q_{ij}$ in stationary regime, integrated over $\tau = \SI{0.2}{\second}$, for different values of $\Delta T$. \textbf{(b)} Average value of $Q_{11}$ and $Q_{12}$ at $\Delta T = \SI{1000}{\kelvin}$ for different integration times $\tau$. For these data $k_1 = \SI{3.6}{\stiffness}$ and $k_2 = \SI{3.7}{\stiffness}$.}
\label{fig:Mean_heat}
\end{figure}

{\it Effective heat fluxes}. 
By analogy with the case of a single trapped Brownian particle~\cite{Sekimoto1998}, we define the \resub{effective} heat dissipated by the particle $i$ during the time interval $\tau$ as:
\begin{equation}
Q_i (\tau) = - \int_{0}^{\tau} \left( - \gamma \dot{x}_i + \xi_i \right) \dot{x}_i \, \mathrm{d}t.
\end{equation}
Using eq.~(\ref{rforce:eq:coupledLangevinsimplified}), we can write $Q_1 = Q_{11} + Q_{12}$ with:
\begin{equation}
\begin{array}{l}
Q_{11} (\tau) = - k_1 \int_{0}^{\tau} x_1 \dot{x}_1 \, \mathrm{d}t \\
Q_{12} (\tau) = - \epsilon k_2 \int_{0}^{\tau} x_2 \dot{x}_1 \, \mathrm{d}t
\end{array}
\end{equation}
and we have the same expressions for $Q_2 = Q_{22} + Q_{21}$ by switching the indexes.
These quantities show several properties expected from real heat fluxes. For example in stationary regime both averages $\langle Q_i \rangle$ are linear in the effective temperature gradient $\Delta T$ and in the integration time $\tau$, as shown in figure~\ref{fig:Mean_heat}. However, these quantities do not exhibit a conservation law, at variance with what is expected in the case of a conservative coupling. Indeed, one can easily show that:
\begin{equation}
\langle Q_1 \rangle = - \frac{k_2}{k_1} \langle Q_2 \rangle
\end{equation}
which means that the total average dissipated heat $\langle Q_1 + Q_2 \rangle$ can be positive or negative depending on the values of $k_1$ and $k_2$ that can be chosen arbitrarily in the experimental set-up.  
However, it has to be pointed out that the energy is not conserved due to the dissipative nature of the coupling forces and not to the artificial random forcing. Indeed the non zero value of the total average dissipated heat corresponds to the  difference of the classical works performed by the dissipative coupling forces. \resub{On the contrary, if $k_1=k_2=k$ the system of equations~(\ref{rforce:eq:coupledLangevinsimplified}) becomes equivalent to one with a conservative coupling:
\begin{equation}
\left\{
  \begin{array}{l}
    \gamma \dot{x}_1 = -\partial{U}/\partial{x_1}  + \xi_1 \\
    \gamma \dot{x}_2 = -\partial{U}/\partial{x_2} + \xi_2
  \end{array}
\right.
\end{equation}\newline
where $U(x_1,x_2)=k/2\times \left(x_1^2 + 2 \epsilon x_1 x_2 + x_2^2 \right)$.
Therefore we retrieve the energy conservation, since $\langle Q_1 + Q_2 \rangle = \left\langle U(x_1(\tau), x_2(\tau))- U(x_1(0), x_2(0)) \right\rangle = 0$ in this case.}

{\it Fluctuation Theorems}. 
In spite of the problems induced by the dissipative coupling, the distribution properties of $Q_i$ show interesting behaviors in stationary regime (when the particle $1$ has already been at $T_1=T+\Delta T$ for a long time), and in transient regime (when the two particles are initially at equilibrium $T_1=T=T_2$ and the effective temperature on particle $1$ is suddenly switched to $T_1=T+\Delta T$ at time $t=0$).

Similarly to the case of conservative forces~\cite{Fogedby12}, by  using the symmetries of the Fokker-Plank operator for the heat Probability Distribution Functions (PDF) in the stationary regime, it can be analytically shown~\cite{NewPap} that $Q_{12}$ verifies an exchange Fluctuation Theorem (xFT) for long integration times $\tau$. In  the limit where $\tau$ tends to infinity, the symmetry function $\Sigma$ verifies:
\begin{equation}
\Sigma(Q_{12}) \equiv \ln \left( \frac{P(Q_{12})}{P(-Q_{12})} \right) = \left( \frac{1}{k_{\text{B}} T_1}-\frac{1}{k_{\text{B}} T_2} \right) Q_{12}
\end{equation}
This xFT is analogous to the one presented in~\cite{Jarzynski2004} for the heat exchanged between two heat bath put in contact during a time $\tau$. Here, the quantity $Q_{21}$ also verifies the same xFT, but corrected by a pre-factor $k_2/k_1$:
\begin{equation}
\Sigma(Q_{21}) \equiv \ln \left( \frac{P(Q_{21})}{P(-Q_{21})} \right) = \frac{k_2}{k_1} \left( \frac{1}{k_{\text{B}} T_2}-\frac{1}{k_{\text{B}} T_1} \right) Q_{21}.
\end{equation}

\begin{figure}[!ht]
\includegraphics[width=0.9\linewidth]{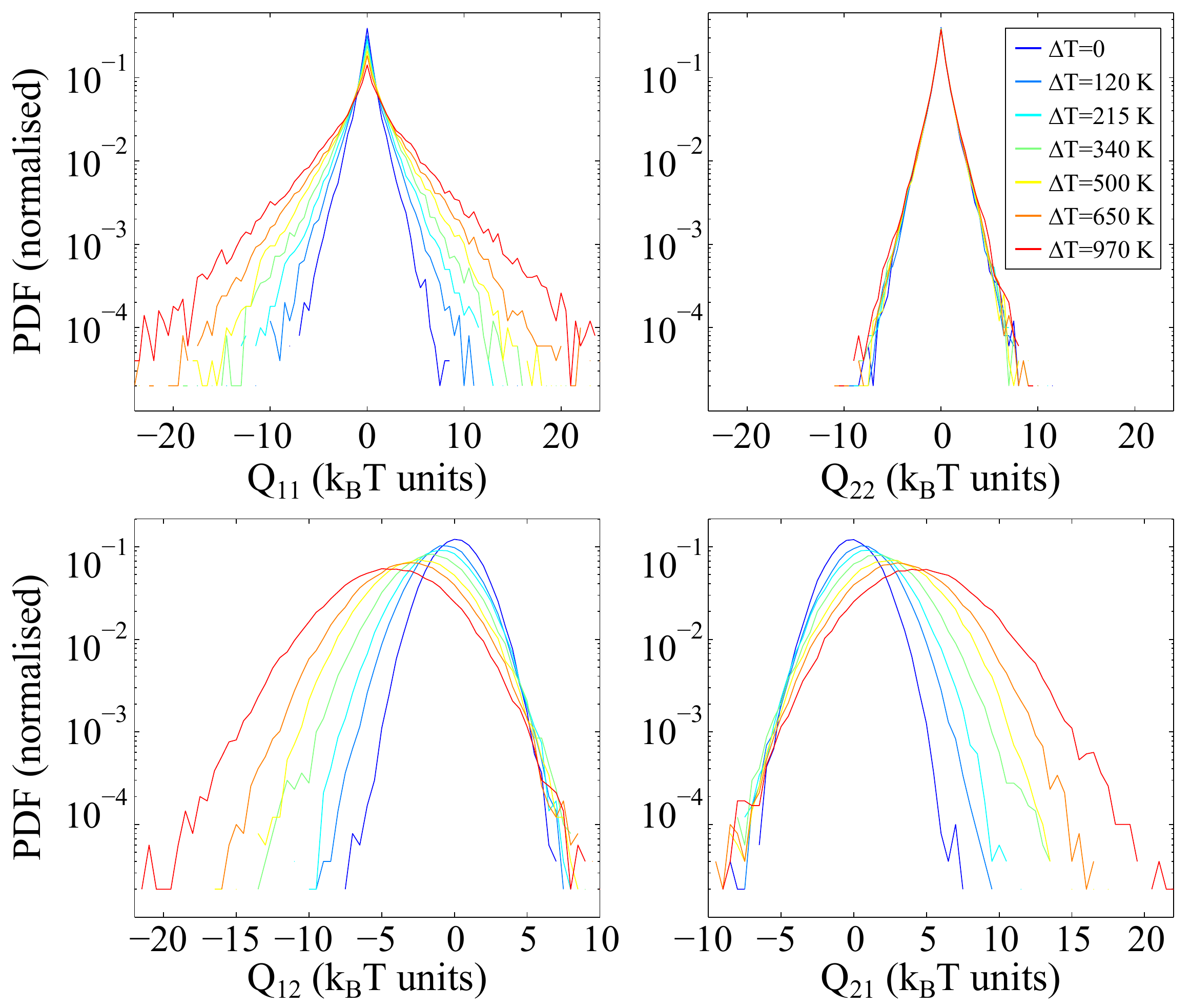}
\caption{Experimental Probability Distribution Functions (PDF) of the four $Q_{ij}$ in stationary regime, integrated over $\tau = \SI{0.25}{\second}$ for different values of $\Delta T$.}
\label{fig:Distribution_Qij}
\end{figure}

\begin{figure}[!ht]
\includegraphics[width=0.9\linewidth]{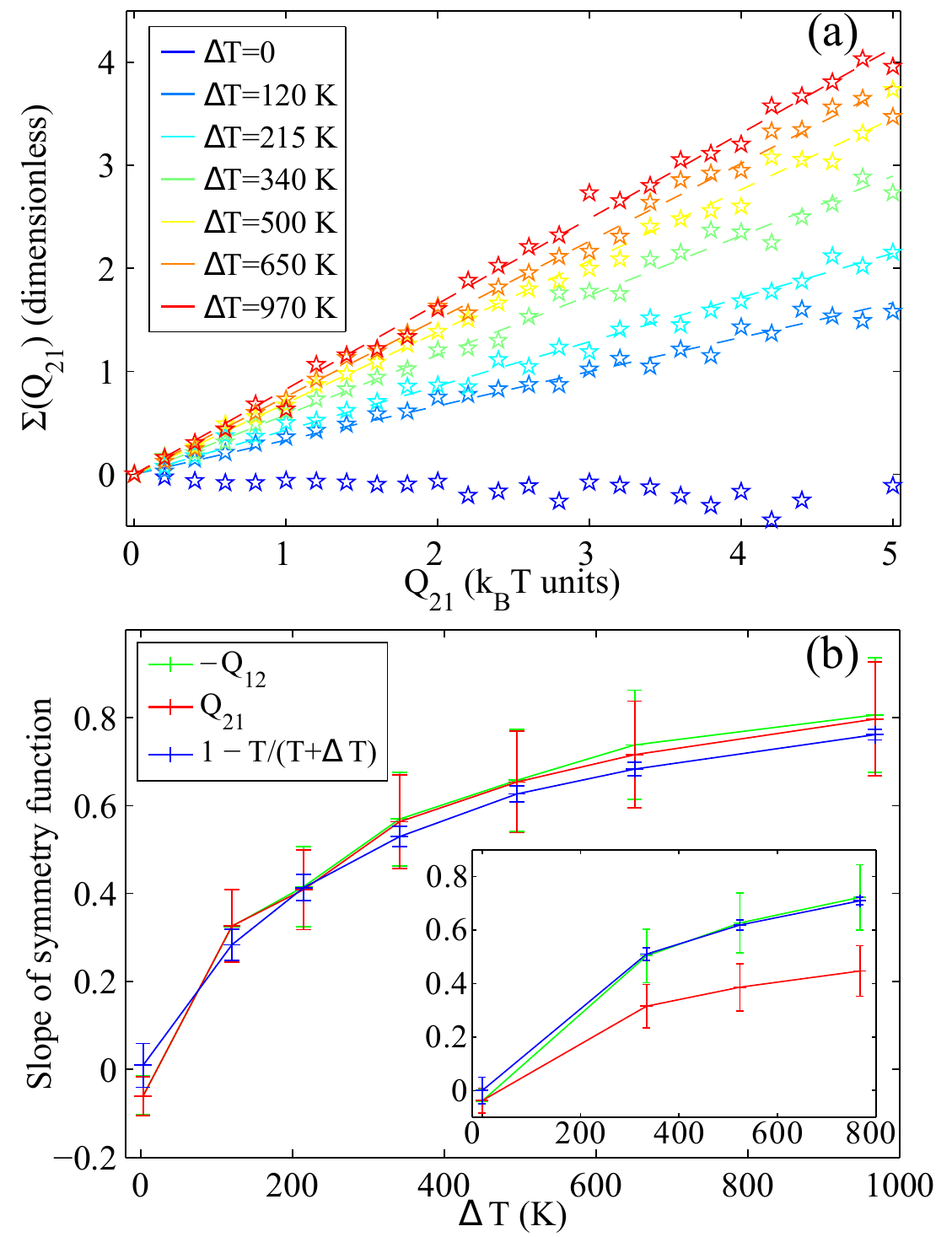}
\caption{\textbf{(a)} Symmetry function $\Sigma (Q_{21})$ as a function of $Q_{21}$ (in \si{\kbT} units) for different values of $\Delta T$, in stationary regime. \textbf{(b)} Slopes of the symmetry function as a function of $\Delta T$, in stationary regime. The main figure is for the case where $k_1 = k_2 = \SI{3.35}{\stiffness}$. The inset is for the case where $k_1=\SI{4.20}{\stiffness}$ and $k_2 =\SI{2.55}{\stiffness}$. The error bars are calculated from the uncertainties on $k_1$, $k_2$ and $\epsilon$.}
\label{fig:Symmetry_functions_stationnary}
\end{figure}

This regime is easy to access experimentally because the typical relaxation time to reach the stationary state is about \SI{0.05}{\second}. The PDF of $Q_{ij}$ integrated over $\tau = \SI{0.25}{\second}$ are shown in figure~\ref{fig:Distribution_Qij} for different values of $\Delta T$.
The symmetry function for $Q_{21}$ is shown in figure~\ref{fig:Symmetry_functions_stationnary}~(a). The linearity of the symmetry function is verified both for $Q_{12}$ and $Q_{21}$ for any value of $\Delta T$. The slopes of the symmetry function are shown in figure~\ref{fig:Symmetry_functions_stationnary}~(b) and also verify the predictions of the xFT, in both cases where $k_1=k_2$ and where $k_1 \neq k_2$. In the case where $k_1 \neq k_2$, the ratio of the slopes of $\Sigma(Q_{21})$ and $\Sigma(Q_{12})$ is equal to the ratio $k_2/k_1$ to a good approximation ($\approx 0.6$ here). Note that since we trace $Q$ in \si{\kbT} units, the expected slope for $\Sigma(Q_{12})$ is simply $T_2/T_1 - 1 = T/(T+\Delta T) - 1$. The value of $\tau$ (\SI{0.25}{\second}) was chosen by computing the PDFs for different $\tau$ to see when it is long enough to have no evolution in the slope of the symmetry function. It is also important to notice that the xFT is satisfied with the value of $\Delta T$ which is the kinetic temperature that can be directly measured from the variance of $x_1$ when the second particle is either not present or at a distance where the coupling is negligible.

In the transient regime, where the system is initially at equilibrium ($T_1=T=T_2$) and the effective temperature is switched on ($T_1=T+\Delta T$) at $t=0$, it has been shown for a conservative system~\cite{Imparato2014} that the heat dissipated by the first bath verifies an xFT for any finite integration time $\tau$. The symmetry function verifies:
\begin{equation}
\Sigma(Q_{1}) \equiv \ln \left( \frac{P(Q_{1})}{P(-Q_{1})} \right) = \left( \frac{1}{k_{\text{B}} T_1}-\frac{1}{k_{\text{B}} T_2} \right) Q_{1},
\label{eq:xFTtransient}
\end{equation}
while $Q_2$ is not supposed to verify such a relation. \resub{Note that we now focus on the total effective heat exchanged $Q_1$, whereas we only considered $Q_{12}$ in the stationary regime. As detailed in~\cite{Imparato2014}, the different behavior between $Q_1$ and $Q_2$ is due to the different initial conditions for each of the two particles: only the temperature of particle 1 is changed at $t=0$.}

\begin{figure}[!ht]
\includegraphics[width=0.9\linewidth]{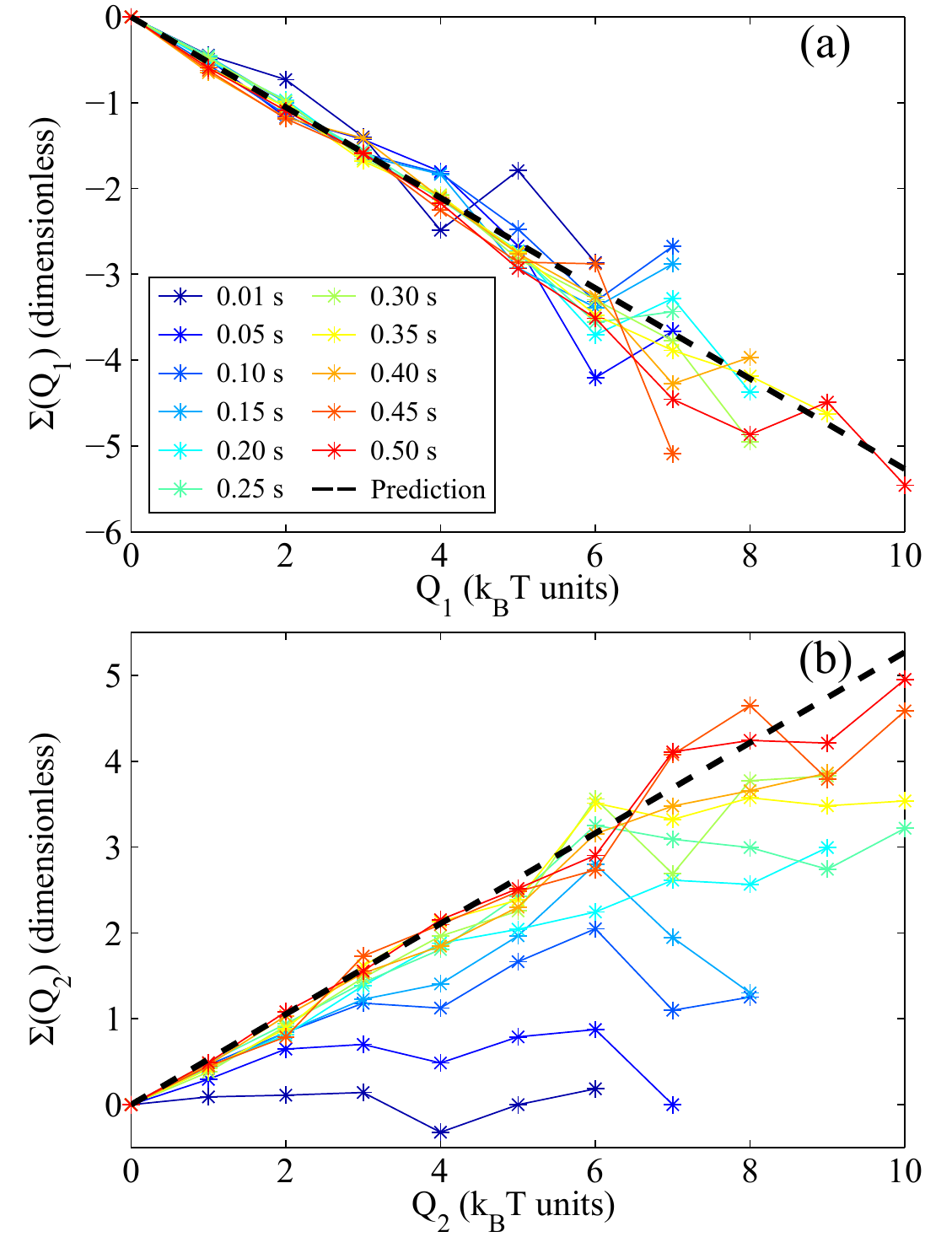}
\caption{\textbf{(a)} Symmetry function $\Sigma (Q_{1})$ as a function of $Q_{1}$ (in \si{\kbT} units) for different values of the integration time $\tau$, in transient regime. The prediction is a linear function of $Q_1$ with a slope equal to $T/(T+\Delta T) - 1$. \textbf{(b)} Symmetry function $\Sigma (Q_{2})$ as a function of $Q_{2}$ (in \si{\kbT} units) for different values of the integration time $\tau$, in transient regime.}
\label{fig:Symmetry_functions_transient}
\end{figure}

The transient regime is experimentally accessible, but requires a very long experimental procedure because each transition from $T_1=T$ to $T_1=T+\Delta T$ only provides one trajectory of $Q_1(\tau)$ and $Q_2(\tau)$, where $\tau=0$ is the time at which $T_1$ has been changed. Furthermore we have to keep the system unperturbed at $\Delta T=0$ for a suitable time interval between two transitions in order to be sure that the  system is at equilibrium before the effective temperature switching. The data shown here are computed for a set of 4375 independent transient regimes, with $\Delta T = \SI{330}{\kelvin}$ and $k_1 \approx k_2 \approx \SI{3.4}{\stiffness}$. The symmetry functions for $Q_1$ and $Q_2$ are shown in figure~\ref{fig:Symmetry_functions_transient}. We see that even if the $\Delta T$ is only a kinematic temperature difference, the equation~(\ref{eq:xFTtransient}) is verified for $Q_1$ for any integration time $\tau$. On the contrary the symmetry function of $Q_2$ exhibits the linear behaviour with the expected slope only for long $\tau$.

{\it Conclusion}.
In this letter we have presented several new results on the energy exchanged between two Brownian particles coupled by viscous interactions and kept at different effective temperatures by an external random forcing on one of the two particles. The effective temperature of the forced particle can be determined by the variances of the particles positions~\cite{Berut2014}. This choice allows us to define an effective heat flux which is a linear function of the temperature difference and which satisfies the stationary exchange fluctuation theorem. Besides we give experimental evidence  that during the transient regime the statistical properties of the heat flowing from the hot to the cold particle are different from those of the heat flowing in the opposite direction, i.e. the first satisfies the transient xFT for any time whereas the second only asymptotically. This interesting and new property has been predicted for systems with a conservative coupling~\cite{Imparato2014}. Here we experimentally prove that it also applies in the case of a viscous coupling, and the theoretical proof will be discussed in a forthcoming publication~\cite{NewPap}. Finally we have shown the difference between the symmetric and asymmetric systems. In the asymmetric case the sum of the total energy fluxes does not satisfy energy conservation. This behavior is only due to the dissipative nature of the coupling and it is not induced by the random forcing. In the perfectly symmetric case the dissipative nature of the coupling cannot be seen and the energy fluxes due to the effective temperature difference behave as real heat fluxes. Indeed theses fluxes not only satisfy the above mentioned statistical properties but also the energy conservation. These results are particularly relevant in all the cases in which an external unknown random forcing is applied to a system which is coupled to another one.

\begin{acknowledgments}
We thank Ken Sekimoto for the very useful discussions we had. This work has been partially supported by the ERC contract OUTEFLUCOP. A.I. is supported by the  Danish Council for Independent Research, and the COST Action MP1209 ``Thermodynamics in the Quantum Regime''.
\end{acknowledgments}

\bibliography{biblio}

\end{document}